\documentclass[journal,twoside,web]{ieeecolor}
\usepackage{generic}
\usepackage{cite}
\usepackage{amsmath,amssymb,amsfonts}
\usepackage{algorithmic}
\usepackage{graphicx}
\usepackage{textcomp}

\usepackage{bm}

\def\BibTeX{{\rm B\kern-.05em{\sc i\kern-.025em b}\kern-.08em
    T\kern-.1667em\lower.7ex\hbox{E}\kern-.125emX}}
\begin{document}
\title{Uniaxial Strain Effects on Graphene Nanoribbon
Resonant Tunneling Transistors}
\author{Mahmood Akbari, Alireza Baghai-Wadji \IEEEmembership{Senior Member, IEEE}, and Razieh Morad
\thanks{Manuscript received //; revised ///; accepted ///. Date of current version ///. The review of this paper was arranged by ///. This work was supported by the Council for Scientific and Industrial Research (CSIR), Pretoria, South Africa}.
\thanks{M. Akbari is with the Department
of Electrical Engineering, University of Cape Town, South Africa (e-mail: makbari@tlabs.ac.za). }
\thanks{A. Baghai-Wadji is with the Department
of Electrical Engineering, University of Cape Town, South Africa (e-mail: alireza.baghai-wadji@uct.ac.za).}
\thanks{R. Morad is with the Department
of Electrical Engineering, University of Cape Town, South Africa (e-mail: rmorad@tlabs.ac.za).}}

\maketitle

\begin{abstract}
The effect of the uniaxial strain on the current-voltage characteristic of a typical armchair graphene-nanoribbon-hBN heterostructure device is simulated numerically by employing the nearest-neighbor tight-binding model and the non-equilibrium Green's function formalism. Simulations clearly reveal the following notable dependencies: (i) the strain invariably reduces the current; (ii) the strain applied in the armchair direction markedly widens the main peak of the current over a larger region of the bias voltage compared with the unstrained state; (iii) the current decreases faster when the strain is applied in the armchair rather than the zigzag direction.
\end{abstract}


\section{Introduction}
\label{sec:introduction}
\IEEEPARstart{G}{raphene} is an atomically thin two-dimensional (2-D) crystal \cite{1 of Intro} with unique thermal, mechanical, and electronic transport properties \cite{2 of Intro}, e.g.,
involving carriers with high mobility, perfect 2-D confinement, and linear dispersion. Consequently, graphene has attracted much attention as a promising candidate for nanodevices over the past decades. Multilayer stacks of graphene and other stable, atomically thin, 2-D materials offer the prospect of creating a new class of heterostructure materials. Hexagonal Boron-Nitride (hBN) is a prominent candidate to be stacked with graphene due to an atomically 2-D layered structure, having a lattice constant very similar to that of graphene (1.8 $\%$ mismatch), comparatively large electrical bandgap (4.7 eV), and excellent thermal and chemical stability \cite{3 of Intro}. The graphene/hBN based tunneling transistors exhibit resonant tunneling, and possess a strong negative differential resistance (NDR) \cite{4 of Intro}- \cite{7 of Intro}. These devices, promising considerable potential for future high-frequency and logic applications, e.g., high-speed IC circuits, signal generators, data storage \cite{8 of Intro}- \cite{10 of Intro} have thoroughly been studied both theoretically and experimentally \cite{11 of Intro}- \cite{17 of Intro}.

In this paper, the effect of the uniaxial strain on graphene-nanoribbon resonant tunneling transistors (RTTs) has been simulated numerically. The uniaxial strain may be induced by either external stress applied to the graphene, in a particular direction, or the substrate on top of which it is deposited. The strain modifies distances between carbon atoms leading to different hopping amplitudes among neighboring sites \cite{Strain0}- \cite{IEEEREF}. The device structure used in the simulations involves an RTT consisting of armchair graphene nanoribbon (AGNR) electrodes with three layers of hBN tunnel barrier sandwiched between them. Employing the nearest-neighbor tight-binding (TB) method and the non-equilibrium Green's function (NEGF) formalism \cite{15 of Intro}- \cite{20 of Intro} the electronic transport characteristics of the RTT is computed. The paper primarily focuses on quantitatively determining the manner strain affects the current-voltage characteristics of AGNR/hBN RTTs.

The paper is organized as follows: In Section II the device structure, the NEGF formalism and the theory of the uniaxial strain are briefly described. Section III is dedicated to the results of the transmission function and the current-voltage characteristics of the device subject to the strain. Conclusions follow in Section IV.

\section{DEVICE GEOMETRY AND METHODS}

\subsection{Device structure}

\begin{figure}[!t]
\centerline{\includegraphics[width=\columnwidth]{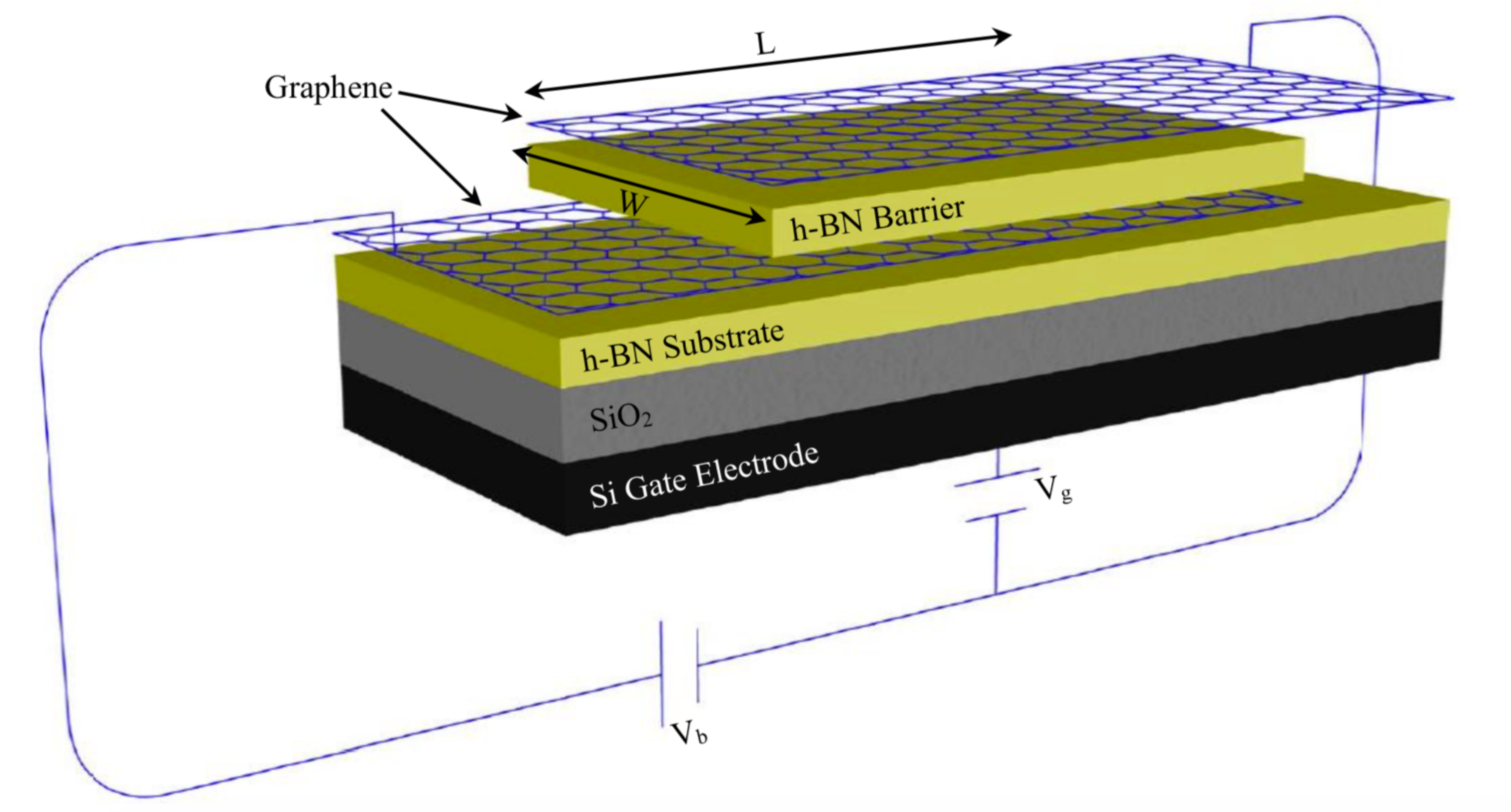}}
\caption{A schematic diagram of the AGNR/hBN RTT. The active region of the device, having length $L$ and width $W$, is placed on a thick hBN substrate, which is positioned on a silicon dioxide substrate. The active region consists of two AGNR electrodes on either side of a tunnel barrier of three hBN atomic layers.}
\label{fig:device}
\end{figure}

The geometry of  AGNR/hBN RTT is illustrated in Fig. \ref{fig:device}. The active region, having length $L$ and width $W$, consists of two semi-infinitely long monolayer armchair-edged graphene nanoribbon (AGNR) electrodes sandwiching three thin hBN layers. 

The bottom AGNR electrode acts as the source, the top AGNR electrode serves as the drain, and the substrate plays the role of the gate electrode of the RTT.
By applying the bias voltage, $V_b$ a tunnel current flows between the source- and the drain electrodes. The current can be resonantly enhanced if the Dirac points of the two AGNR electrodes are aligned \cite{4 of Intro} by tuning the gate voltage $V_g$. 

The armchair electrodes considered in this device are metallic with  even number of the ribbon width index $N$ = 6$p$ + 2 ($p$ = 1,2,3,$\ldots$) and nanoribbon width of $W$ = $(N-1) a_0/2$. Here,  $a_0$ stands for the graphene lattice constant ($a_0$ = 0.246 nm) as shown in Fig. \ref{fig:geometry}. It is assumed that the layers have an AB order (Bernal stacking) \cite{23 of Intro}- \cite{28 of Intro}. The lattice constant mismatch between hBN and graphene (1.8 $\%$) \cite{13 of Intro}, has been neglected.

Since the metallic graphene layers have much higher conductivity than hBN, the applied bias voltage rigidly shifts the electrostatic potential energy of the bottom graphene electrode by the amount equal to $U=-eV_b$, while the value of the electrostatic potential energy of the top graphene layer remains zero.
The electrostatic potential of the hBN layers, sandwiched between the bottom- and the top electrodes, are obtained by assuming a linear potential profile between the electrodes. The bias voltage controls the chemical potential of contacts $\mu_B = -eV_b$ and $\mu_T$ = 0 for the bottom and top graphene leads, respectively. The gate voltage shifts the electrostatic potential energy at the bottom electrode by $\Delta U$ = - 0.01 e$V_g$ \cite{7 of Intro}.

\subsection{NEGF Formalism}
The non-equilibrium Green's function (NEGF) formalism \cite{33 of Intro} has been employed in this simulations. The general retarded Green's function is computed by

\begin{equation}\mathbf{G}^r(E)=[E\, \mathbf{I} -\mathbf{H}-\mathbf{\Sigma_L^r}(E)-\mathbf{\Sigma_R^r}(E)]^{-1}\, ,\label{eq:broadeningf}\end{equation}
where $E$, $\mathbf{I}$, and $\mathbf{H}$ are the energy of electrons, the identity matrix, and the Hamiltonian matrix of the system, respectively. Furthermore, the $\mathbf{\Sigma^r_{L/R}}$ is the self-energy of the semi-infinite left (right) graphene layer (refer to Fig. \ref{fig:device}). $\mathbf{\Gamma_{L/R}}(E)$ are evaluated using self-energies given by,

\begin{equation} 
\mathbf{\Gamma_{L}}(E)= i\, [\mathbf{\Sigma_L^r}(E)-\mathbf{\Sigma_L^r}^{\dagger}(E)], \label{eq:broadeningf1}
\end{equation}

\begin{equation}
\mathbf{\Gamma_{R}}(E)= i\, [\mathbf{\Sigma_R^r}(E)-\mathbf{\Sigma_R^r}^{\dagger}(E)], \label{eq:broadeningf2}
 \end{equation}
where the dagger signifies the Hermitian conjugate. 

The transmission function can be calculated using the expression

\begin{equation}T(E)= Tr [\mathbf{\Gamma_{L}}(E)\, \mathbf{G}^r(E)\, \mathbf{\Gamma_{R}}(E)\,\mathbf{G}^{r\dagger}(E)]. \label{eq:transmission}\end{equation}

The current as a function of bias voltage is

\begin{equation}I= \frac{2e}{h}\, \int \,\,T(E)\, [f(E-\mu_R)-f(E-\mu_L)]\,dE,\label{eq:IV}\end{equation}
where $f(E-\mu_{L/R})$ is the Fermi-Dirac distribution function for the left (top) or right (bottom) graphene layers.

The recursive Green's function (RGF) approach has been adopted to calculate the retarded Green's function \cite{15 of Intro}. Additionally, the Sancho iteration method \cite{Sancho} has been used to evaluate the self-energy of the leads.

\subsection{Uniaxial Strain}
When graphene is deposited on substrates, structural deformation arises naturally due to distinct atomic arrangements of graphene and the substrate. Theoretical calculations \cite{69} and experiments \cite{70} have shown that strain can change the initial inter-atomic distance up to  20$\%$ without opening a band gap.

In this work, the graphene layers are considered uniformly stretched or compressed along a prescribed direction $\theta$ with respect to the Y-direction in the graphene plane. In addition, since three hBN layers have been considered between the graphene electrodes, the same strain is assumed to be applied to the hBN layers as well. 

\begin{figure}[!t]
\centerline{\includegraphics[width=\columnwidth]{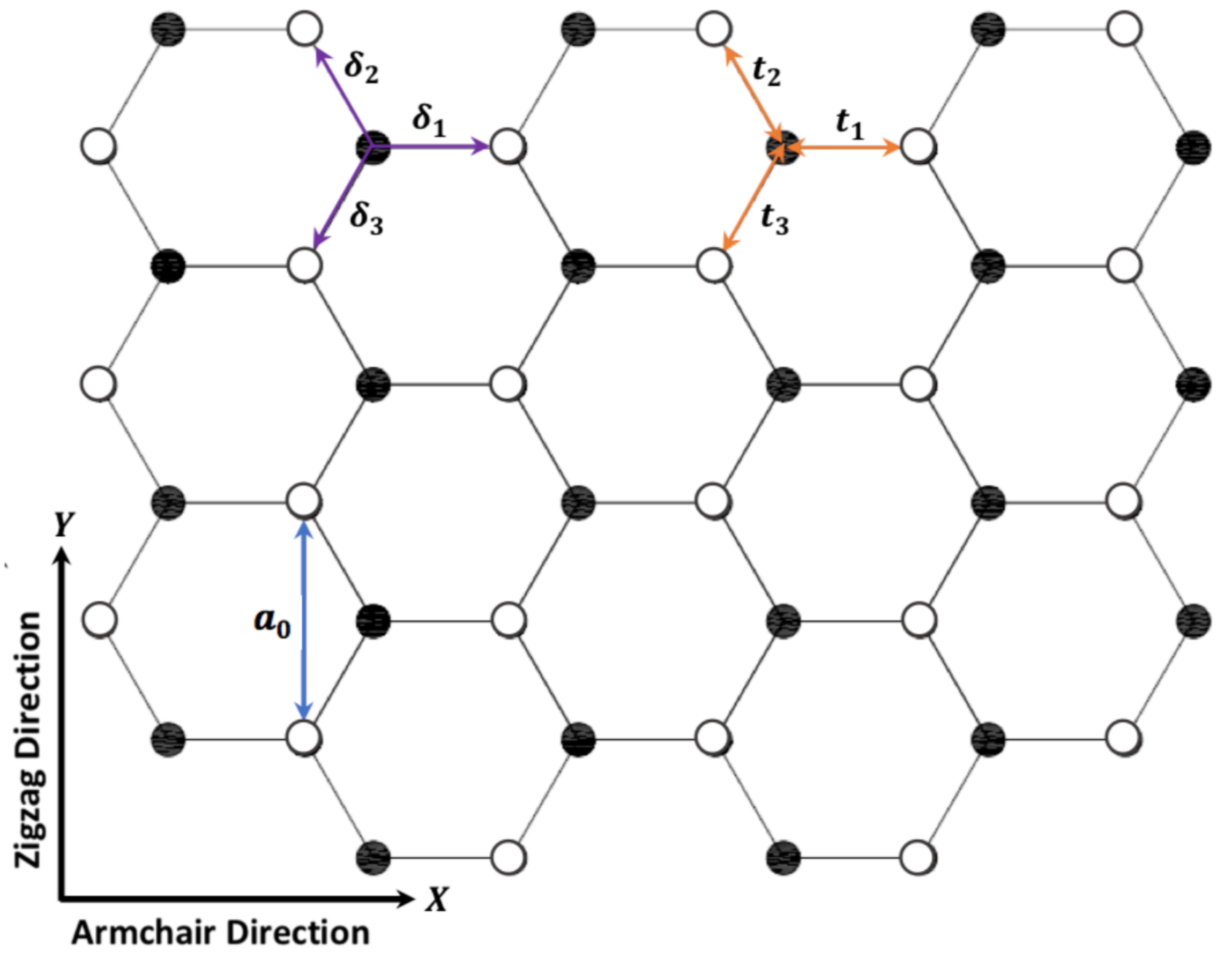}}
\caption{Schematic geometry of an armchair nanoribbon. The parameter $a_0$ is the lattice constant. The X and Y directions, respectively, are along the armchair and zigzag edges. Here, $\bm{\delta_i}$ is the distance between the nearest-neighbors, while $t_i$ characterizes the hopping between the nearest sites. By applying strain, the distances $\bm{\delta_i}$ are changed depends on the direction of the strain, resulting in a modification of the hopping parameters.}
\label{fig:geometry}
\end{figure}

Considering the Y-direction along the zigzag direction of the lattice in Fig. \ref{fig:geometry}, the strain tensor in the lattice coordinate system is \cite{Strain1}
\begin{equation}
  \bm{\epsilon} = \ensuremath{\epsilon_0}  \begin{pmatrix}
     \cos^2\theta -\sigma\sin^2\theta & 
     (1+\sigma)\cos\theta\sin\theta \\
     (1+\sigma)\cos\theta\sin\theta &
     \sin^2\theta -\sigma\cos^2\theta
  \end{pmatrix}
  \,,
  \label{eq:StrainTensor}
\end{equation}
where $\ensuremath{\epsilon_0} $ is strain modulus, $\theta$ is the direction of the applied strain, and $\sigma$ is the Poisson ratio, which is equal to 0.165 for graphene \cite{Strain0}- \cite{Strain1}. 

The distance between the atoms in each layer after the strain application, needs to be adjusted according to
\begin{equation}
\bm{\delta^s}=(\bm{1}+\bm{\ensuremath{\epsilon}})\cdot\bm{\delta} , 
\label{eq:Gen-Vec-Deformation}
\end{equation}
which leads to distortion of the reciprocal lattice. Here, $\mathbf{\delta}$ stands for the distance between the nearest-neighbors (refer to Fig. \ref{fig:geometry}) while the $\mathbf{\delta^s}$ refers to the nearest-neighbors distance under the strain. The change in bond lengths leads to different hopping amplitudes among intralayer neighboring sites. It is generally accepted that the following exponential relationship models the hopping parameter of graphene under strain sufficiently accurately \cite{Strain0}- \cite{Strain2}

\begin{equation}
 t^s = t \, e^{-3.37(|  \bm{\delta^s} |/|  \bm{\delta} |-1)},
  \label{eq:Hopping}
\end{equation}
where $t$ represents the intralayer hopping parameters shown in Table \ref{table}. The superscript $s$ refers to the strain. The rate of decay is extracted from the experimental results for the graphene \cite{CastroNeto:2007b}. The direction of the strain is defined such that $\theta$= 0 corresponds to the zigzag direction of the lattice while $\theta= \pi/2$ represents the armchair direction.

\section{RESULTS}
\begin{figure}[!t]
\centerline{\includegraphics[width=\columnwidth]{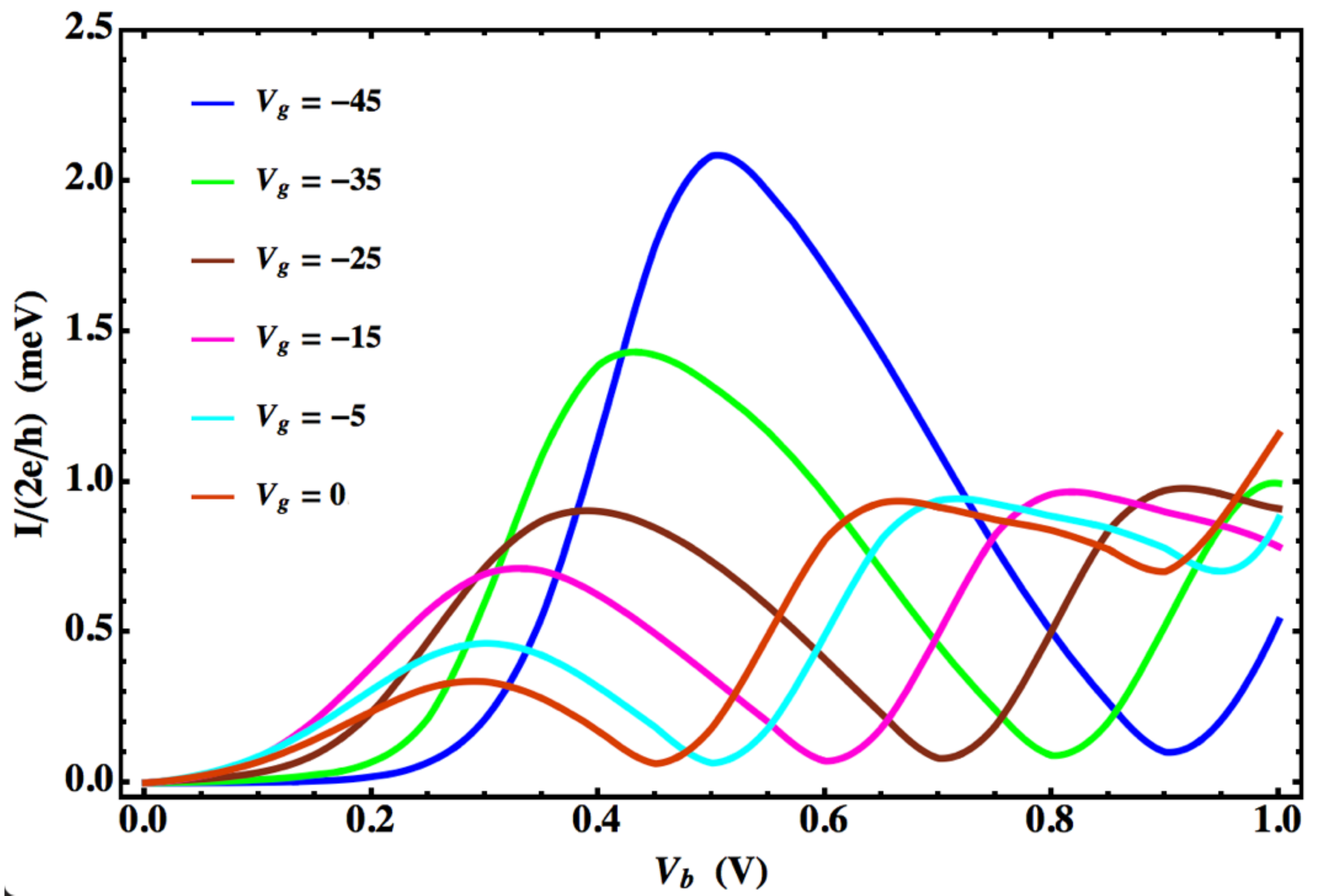}}
\caption{Current-Voltage characteristics as a function of bias voltage $V_{b}$ at fixed gate voltages at $T$ = 300 K for AGNR/$\mathrm{(hBN)_{3}}$/AGNR resonant tunneling transistor with $W$ = 7.6 nm and $L$ = 6.8 nm.}
\label{I-V-unstrained}
\end{figure}

Figure. \ref{I-V-unstrained} shows the current-voltage characteristics as a function of bias voltage $V_{b}$ at fixed gate voltages $V_{g}$ for the AGNR$/ \mathrm{(hBN)_{3}}$/AGNR resonant tunneling transistor with $N$ = 62 ($W$ = 7.6 nm) and $M$ = 16 ($L$ = 6.8 nm) at $T$ = 300 K, with $N$ being the number of dimer lines and $M$ the number of unit cells. The Table \ref{table} represents the tight-binding parameters employed in the simulations \cite{23 of Intro}, \cite{25 of Intro}. The inter-layer distance between the AGNR and hBN layers is assumed to be 0.36 nm, the inter-layer distance between the hBN layers is 0.33 nm, and the relative dielectric constant of the insulator is taken to be 3.9.

\begin{table}
\caption{The nearest-neighbor tight-binding parameters used in this paper \cite{23 of Intro}, \cite{25 of Intro}. }
\label{table}
\setlength{\tabcolsep}{3pt}
\begin{tabular}{|p{75pt}|p{75pt}|p{75pt}|}
\hline
On-site Energies / eV          & 
Intra-layer Hopping / eV       & 
Inter-layer Hopping / eV \\
\hline
$E_C$= 0              & 
$t_{C-C}$= 2.64      & 
$t_{B-N}$= 0.6   \\
$E_B$= 3.34           & 
$t_{B-N}$= 2.79       & 
$t_{C-B}$= 0.43  \\
$E_N$= -1.4           & 
                       & 
$t_{C-N}$= 0.43  \\
\hline
\end{tabular}
\end{table}

In this study, the focus is on the effect of uniaxial strain on the current-voltage characteristics of the device depicted in Fig. \ref {fig:device}. It is assumed that the uniaxial strain applied to the multilayer device is the same in each layer. Consequently, all graphene and hBN layers are under the same tension, and, it is further assumed that the tension does not affect the distance between inter-layer atoms and inter-layer hopping parameters. Figure. (\ref {fig:geometry}) shows the schematic geometry of armchair nanoribbon. The Y-direction denotes the zigzag edge, while the X-direction indicates the armchair edge. The distance between the nearest-neighbors is given by $\bm{\delta}$, and $t$ characterizes the hopping parameter between nearest sites. By applying strain, these distances change depending on the direction of the tension which in turn modifies the hopping parameters (Refer to Section II.C for more details) and the position of the Dirac points. 

\begin{figure}[!t]
\centerline{\includegraphics[width=\columnwidth]{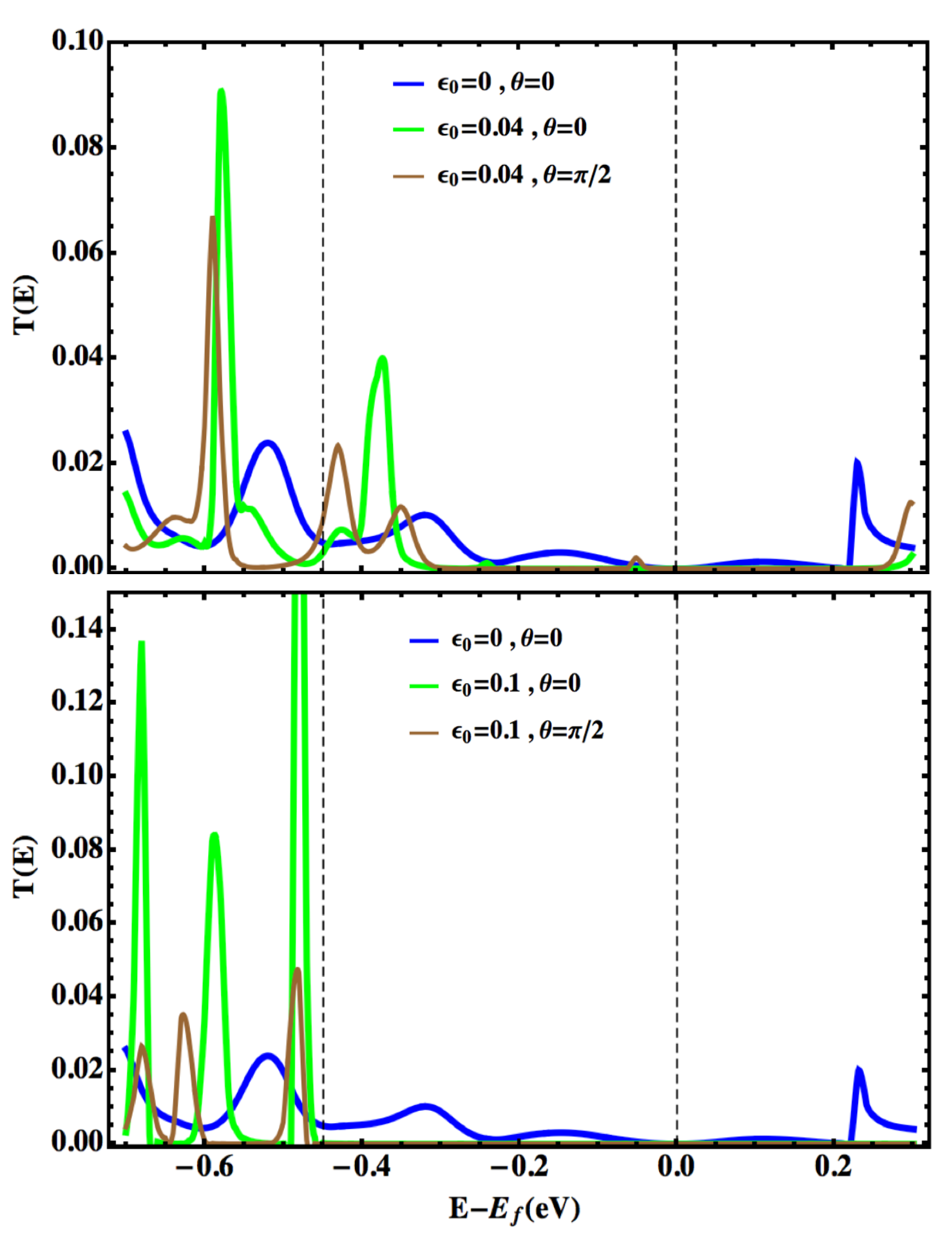}}
\caption{The transmission function as a function of energy. The dashed black lines show the energy window determined by the Dirac points of two electrodes for two strain modulus applied to both the zigzag and armchair directions for $V_{g}$ = -45 V and $V_{b}$ = 0.45 V. Applied strain renders the peaks sharper in the active region.}
\label{fig:tr}
\end{figure}

In Fig. \ref{fig:tr}, the transmission function is plotted for different strain modulus applied to both the zigzag and armchair directions for $V_g$ = -45 V and $V_b$ = 0.45 V. Vertical dash-dot lines in Fig. \ref{fig:tr} give the chemical potential at both AGNR ends $\mu_B$ and $\mu_T$, which determines the bias window. The current can be calculated by integrating the transmission function in bias window, \eqref{eq:IV}, such that the transmission peaks in this window mainly contribute to the current. All sharp peaks represent the tunneling peaks due to the energy alignment of subbands in the top and bottom graphene contacts; they do not contribute significantly to the current. The transmission functions show that the electrons at $E$ = 0 eV encounter a strong barrier between the AGNR layers. For comparatively small strain values $\ensuremath{\epsilon_0}$ = 0.04 in zigzag direction ($\theta$ = 0) the main peak in Fig. \ref{fig:tr}(top) is increasing and becoming sharper which leads to a smaller current. The behavior of the transmission function for the applied strain in the armchair direction ($\theta$ = $\pi$/2)  is in principle the same: peaks are increasing and becoming sharper. By increasing the strain modulus (See Fig. \ref{fig:tr}(bottom) for $\ensuremath{\epsilon_0}$ = 0.1) the main peaks in the active energy regions in both directions collapse.

\begin{figure}[!t]
\centerline{\includegraphics[width=\columnwidth]{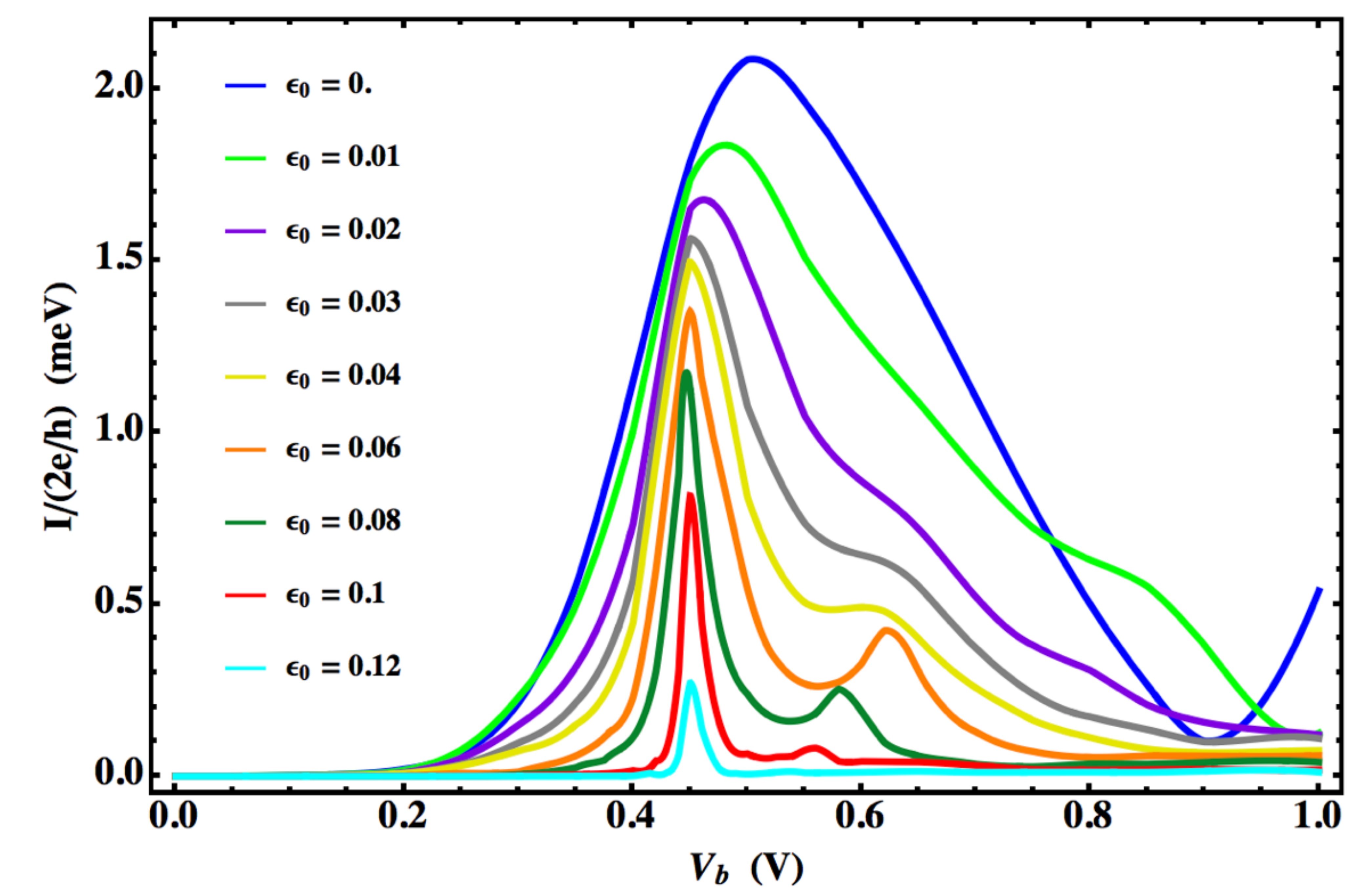}}
\caption{The current-voltage characteristic of the device under the uniaxial strain in zigzag direction($\theta$ = 0). The gate voltage is $V_{g}$= -45 V. Increasing the strain modulus causes a decrease in the current.}
\label{fig:IVZ}
\end{figure}

\begin{figure}[!t]
\centerline{\includegraphics[width=\columnwidth]{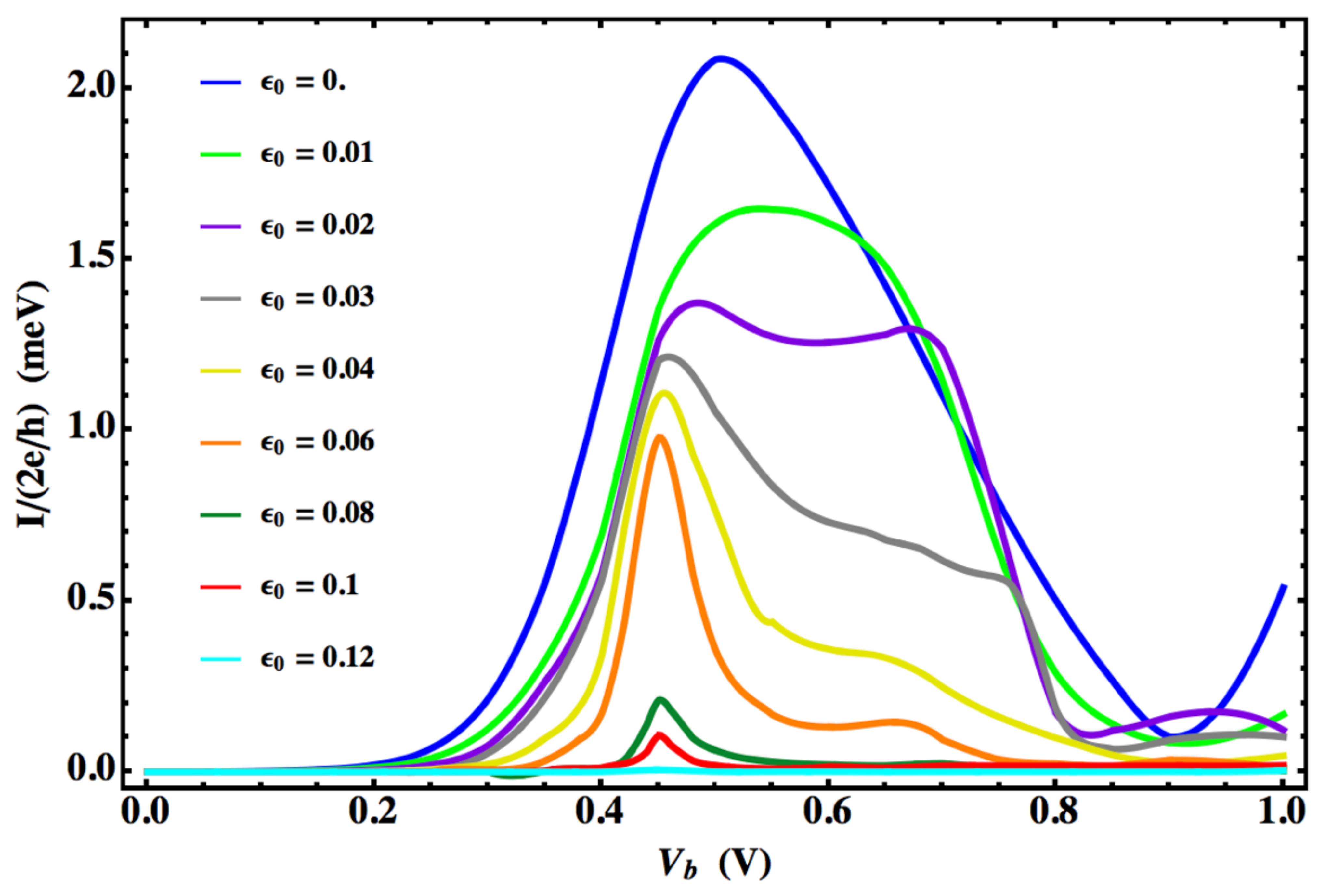}}
\caption{The current-voltage characteristic of the device under the uniaxial strain in armchair direction ($\theta$ = $\pi$/2). The gate voltage is $V_{g}$ = -45 V. Increasing the strain modulus causes decreasing the current.}
\label{fig:IVA}
\end{figure}

From a physical point of view, the application of strain causes a change in the interaction between graphene atoms, such that the increase of the strain modulus leads to a stronger localization of the $P_{z}$ orbitals of the atoms in the honeycomb lattice; this disrupts the alignment of the Dirac points in the graphene layers during the tunneling process, which results in the collapse of the current. Consequences of these trends can be viewed in the $I-V$ curves. The resultant $I-V$ characteristic of the device for various strain modulus applied to the zigzag direction, $\theta$ = 0 and the armchair direction, $\theta$ = $\pi$/2 are plotted in Fig. \ref{fig:IVZ} and Fig. \ref{fig:IVA}, respectively. The simulations show that the applied strain in the $\theta$ = 0 direction results in the narrowing of the main peak of current, while simultaneously decreasing the maximum amount of the current. In addition, the next smaller peak approaches the main peak and rapidly collapse to zero. Applying strain in the armchair direction, widens the main peak of current over a bigger region in the bias voltage domain, although the maximum of the current continues to decreases. The current decreases faster when the same strain applied in the armchair direction compared with being applied in the zigzag direction. The computed numerical data are consistent with the results reported in \cite{strain3}-\cite{strain5} where it is demonstrated that the electronic structure of strained GNRs strongly depends on its edge shape and the structural indices such that zigzag GNRs are not particularly sensitive to uniaxial strain, while the electronic structure of AGNRs is very sensitive to the strain and even a band gap could appear for large ratio of the deformation.

\section{CONCLUSION}

The effect of uniaxial strain on the I-V characteristic of an AGNR-hBN-AGNR multilayer heterostructure device was studied theoretically.  
This class of devices supports resonant tunneling if the applied gate voltage aligns the Dirac points of the top and bottom graphene electrodes. Applying strain on the device deforms the Dirac points of the graphene electrodes affecting the transport properties of the device. In this work, a uniaxial strain was applied in the armchair- and zigzag directions of the nanoribbons, assuming a uniform change in the intra-layer atomic distances in all layers. For simplicity, it was assumed that inter-layer distances and consequently inter-layer hopping parameters are not all affected by the tension. By using the tight-binding model and non-equilibrium Green's function formalism, the current-voltage characteristic of the device was calculated in the presence of the uniaxial strain. The results demonstrated that the strain decreased the amount of the current in both cases. however, the quantitative behavior of the I-V characteristic turned out to be different for the strain being applied in different directions. The observed collapse of the current occurred more rapidly when the strain was applied in the armchair direction.

The effect of the uniaxial strain on the chosen device enables the control of the current by applying tension to the layers to obtain (design) the desired I-V characteristics. This property may offer attractive applications in NEMS devices. Potential applications in sensing devices also offer themselves. 

The preliminary results obtained in this work suggest that by considering the effect of inter-layer hopping parameters, applying different tensions in various directions (specified by $\theta$), and comparing them with the first principle calculations, will provide deeper insights and valuable guidelines for future theoretical and experimental investigations of strain effect on heterostructure devices.

\section*{ACKNOWLEDGEMENTS}

The authors would like to thank Dr. Mohsen Modarresi, Prof. Nobuya Mori and Dr. Yunqi Zhao for useful discussions. Support from the Council for Scientific and Industrial Research (CSIR), Pretoria, South Africa, is gratefully acknowledged.

\bibliographystyle{IEEEtran}
\bibliography{StrainTransistor}

\end{document}